\documentclass[prb,twocolumn,showpacs,superscriptaddress]{revtex4-1}

\usepackage{bm}
\usepackage{graphicx}
\usepackage{amsmath}

\newcommand{\ket}[1]{\left| #1 \right\rangle}
\newcommand{\bra}[1]{\left\langle #1 \right|}
\newcommand{\braket}[2]{\left\langle #1 | #2 \right\rangle}

\begin{document}
\title{Lattice Density-Functional Theory for Quantum Chemistry}

\author{J. P. Coe}\email{J.Coe@hw.ac.uk}
\affiliation{Institute of Chemical Sciences, School of Engineering and Physical Sciences, Heriot-Watt University, Edinburgh, EH14 4AS, UK.}

\begin{abstract}
We propose a lattice density-functional theory for {\it ab initio} quantum chemistry or physics as a route to an efficient approach that approximates the full configuration interaction energy and orbital occupations for molecules with strongly-correlated electrons. We build on lattice density-functional theory for the Hubbard model by deriving Kohn-Sham equations for a reduced then full quantum chemistry Hamiltonian, and demonstrate the method on the potential energy curves for the challenging problem of modelling elongating bonds in a linear chain of six hydrogen atoms. Here the accuracy of the Bethe-ansatz local-density approximation is tested for this quantum chemistry system and we find that, despite this approximate functional being designed for the Hubbard model, the shapes of the potential curves generally agree with the full configuration interaction results. Although there is a discrepancy for very stretched bonds, this is lower than when using standard density-functional theory with the local-density approximation.
\end{abstract}


\maketitle

\section{Introduction}

Efficient computational methods based on small corrections to a single determinant of one-electron orbitals in {\it ab initio} quantum chemistry and physics can give qualitatively incorrect results if applied to electronically excited states, molecules containing transition metals, or bond breaking. For example when elegant approximations in coupled cluster theory are used to model the dissociation of the nitrogen dimer.\cite{RevModPhysBartlett07} Such problems may require multiple determinants as the starting point of a now computationally intensive calculation and are often termed multireference or even strongly correlated. Full configuration interaction (FCI) gives the most accurate result for a given basis set of one-electron orbitals. However, as the number of determinants scales factorially with the size of the basis set, it is computationally prohibitive for all but the smallest systems. If there are $K$ basis functions and $N_{\uparrow}+N_{\downarrow}$ electrons then the FCI wavefunction will consist of $\binom{K}{N_{\uparrow}} \binom{K}{N_{\downarrow}}$ determinants when there are no symmetries to exploit. This means that for only 20 basis functions and 20 electrons with equal numbers of both spins then there are already $\sim 10^{10}$ configurations and finding their coefficients in the FCI wavefunction by diagonalization of the Hamiltonian matrix will be computationally intractable. By using the electron density rather than the many-electron wavefunction, density-functional theory (DFT) can in principle efficiently describe even strongly-correlated systems. However in practice standard approximate functionals can perform poorly when confronted with multireference problems, e.g., the dissociation of the hydrogen molecule\cite{Cohen08} or spin gaps in transition metal complexes.\cite{Reiher02}

Yet lattice DFT (L-DFT)\cite{LDFT1,LDFT2,LDFT3} with, e.g., the Bethe-ansatz local-density approximation (BA-LDA)\cite{LimaPRL2003} allows strongly-correlated lattice systems, such as Hubbard models, to be successfully and efficiently modelled\cite{Capelle13} where the lattice density or site occupation takes the role of the density in standard DFT. Applications to Hubbard models have included modelling the system with periodic modulations in the external potential and onsite repulsion,\cite{Silva05} using a harmonic potential to investigate the Luther-Emery phase,\cite{Xianlong07} modelling ultracold repulsive fermions in one dimensional optical lattices\cite{Xianlong08} then simulating experimental data of cold atoms in optical lattices,\cite{Campo07} and calculating the site entanglement when external potentials are used.\cite{Franca08} Accurate results for the local density and magnetization in spatially inhomogeneous spin-polarized systems have been obtained using an analytic parametrization for a Bethe-ansatz local-spin-density approximation.\cite{BALSDA} An adiabatic BA-LDA for time-dependent L-DFT has been created\cite{Verdozzi08} and used to model the Coulomb blockade in quantum dots.\cite{Kurth10} The $\mu$-BALDA has been created which uses the local chemical potentials to enable convergence in L-DFT when site densities are close to one.\cite{Ying14} L-DFT for the Hubbard model has been built upon with a one-electron reduced density matrix functional for the interaction energy\cite{Saubanere16} and the iBALDA was developed \cite{Senjean18} for site-occupation embedding theory. The accuracy of approximations in L-DFT has also been appraised using metric space approaches.\cite{LDFT18} Similarly to the Hohenberg-Kohn theorems\cite{HK} providing the foundation of standard DFT, L-DFT depends on the result that the site density uniquely determines the wavefunction\cite{PhysRevB.52.2504,WU1,Godby95} and it was recently proven that for lattice systems, with certain caveats, the wavefunction uniquely determines the external potential.\cite{CoeEPL15}

 The quantum chemistry Hamiltonian in a basis set of single-particle orbitals may be mapped to a lattice system where sites represent orbitals when using the notation of second quantization, which the Hubbard model approximates. This has enabled the impressive use of the powerful approach of the density matrix renormalization group (DMRG)\cite{DMRGWhite92} for multireference problems in quantum chemistry.\cite{ChanDMRGreview2011} A DMRG calculation is systematically improvable by increasing the number of states ($M$), variational and can be successful on molecules that are beyond other wavefunction methods.\cite{Kurashige13} However the accuracy can depend on the orbital ordering when mapped to a lattice system, and for $K$ basis orbitals the scaling\cite{ChanDMRGreview2011} of the calculation as $O(M^{3}K^{3})+O(M^{2}K^{4})$ means it can become computationally intractable as the system size increases if a large $M$ is necessary for accurate results. 

We also consider the quantum chemistry Hamiltonian mapped to a lattice system when using a basis set and use this to create a quantum chemistry lattice DFT (QC-LDFT) to approximate the FCI site density (orbital occupation) and energy as an essential step towards an efficient approach to model multireference problems in quantum chemistry. Although the accuracy will depend on the choice of approximate functional, only site densities from self-consistently solving a non-interacting system are required rather than a many-electron wavefunction for the interacting system. This means that for $K$ basis orbitals one only has to diagonalize $K\times K$ symmetric matrices thereby incurring a computational cost that scales as $O(K^{3})$.\cite{Demmel97}  

In this paper we first briefly discuss L-DFT and the BA-LDA for the Hubbard model. Next we derive the equations for QC-LDFT where we consider a reduced Hamiltonian with interaction terms limited to those involving intersite densities before considering the full quantum chemistry Hamiltonian. The numerical procedures we employ to implement QC-LDFT are then presented. We then go beyond the Hubbard model and demonstrate QC-LDFT for the first time on the potential energy curve for a linear chain of six hydrogens when using 12 basis functions where we test the accuracy of the BA-LDA functional.\cite{LimaPRL2003} We first use the reduced Hamiltonian before considering the full quantum chemistry Hamiltonian for this initial application of QC-LDFT. Despite using an approximate functional designed for the Hubbard model (the BA-LDA)\cite{LimaPRL2003} we capture the shapes of the FCI potential curves and, for the full Hamiltonian, improve upon standard DFT results at stretched bond lengths.

\section{Methods}
\subsection{Lattice Density-Functional Theory}
L-DFT allows the efficient modelling of the Hubbard model when inhomogeneity is introduced through an external potential $v_{ext,i}$. The Hamiltonian of this interacting system is
\begin{eqnarray}
\nonumber \hat{H}_{HM}&=&-t\sum_{i,\sigma} \left( \hat{a}_{i,\sigma}^{\dagger} \hat{a}_{i+1,\sigma}+\hat{a}_{i+1,\sigma}^{\dagger} \hat{a}_{i,\sigma} \right)\\
 &+&U\sum_{i} \hat{n}_{i,\uparrow} \hat{n}_{i,\downarrow}+\sum_{i} v_{ext,i}\hat{n}_{i}
\label{eq:HubbardModelwithVext}
\end{eqnarray}
where $\hat{a}_{i,\sigma}^{\dagger}$ creates a particle of spin $\sigma$ at site $i$ while $\hat{a}_{i,\sigma}$ annihilates it and the number operator or site density operator
is $\hat{n}_{i,\sigma}=\hat{a}_{i,\sigma}^{\dagger} \hat{a}_{i,\sigma}$.

In the Kohn-Sham (KS)\cite{KS} approach to L-DFT, see e.g. Ref.~\onlinecite{Capelle13}, the exact energy is written as a functional of the site density or occupation
\begin{equation}
E[n]=T_{NI}[n]+E_{H}[n]+\sum_{i} v_{ext,i}n_{i} +E_{xc}[n]
\end{equation}
where $T_{NI}$ is labelled the kinetic energy term for non-interacting electrons, $E_{H}$ the Hartree term, and $E_{xc}$ the exchange-correlation energy functional. As the exact form for this latter quantity is unknown then approximations have to be used in practice and therefore the energy is approximate.
The site density or occupation $n_{i}$ is then found by self-consistently solving the non-interacting KS Hamiltonian 
\begin{equation}
\hat{H}_{KS}=\hat{T}+\sum_{i} v_{eff,i}[n] \hat{n}_{i}
\end{equation}
where $\hat{T}=-t\sum_{i,\sigma} \left( \hat{a}_{i,\sigma}^{\dagger} \hat{a}_{i+1,\sigma}+\hat{a}_{i+1,\sigma}^{\dagger} \hat{a}_{i,\sigma} \right)$ is labelled as the kinetic energy operator for the Hubbard model and
\begin{eqnarray}
\nonumber v_{eff,i}[n]&=&\frac{\partial E_{H}}{\partial n_{i}}+v_{ext,i}+\frac{\partial E_{xc}}{\partial n_{i}} \\
&=&v_{H,i}[n]+v_{ext,i}+v_{xc,i}[n].
\end{eqnarray}
 For $N$ electrons when the spins are balanced, the first $N/2$ eigenfunctions $f_{j}$ form the single Slater determinant which gives the site density as
$n=2\sum_{j}^{N/2} |f_{j}|^{2}$ and $T_{NI}=2\sum_{j}^{N/2} \bra{f_{j}} \hat{T} \ket{f_{j}}$ allowing $E[n]$ to be computed.  We emphasize that although the occupation of the $f_{j}$ cannot be fractional, as these are the Kohn-Sham orbitals, the density or occupation at a site is a continuous variable that would be identical to that of the interacting system if the exact exchange-correlation potential $v_{xc}$ were known.

\subsection{Bethe-Ansatz Local-Density Approximation}
A LDA using the Bethe-Ansatz was first introduced for L-DFT in Ref.~\onlinecite{PhysRevB.52.2504}, later the BA-LDA\cite{LimaPRL2003} became a popular and successful approximation to $E_{xc}$ in L-DFT. The BA-LDA interpolates three limiting cases ($U\rightarrow \infty$ with $n \leq 1$ , $U=0$ with $n \leq 1$, and $n=1$) of the exact Bethe-ansatz 
energy results for the homogeneous Hubbard model, i.e, Eq.~\ref{eq:HubbardModelwithVext} when $v_{ext}=0$. The interpolation uses\cite{LimaPRL2003} 
\begin{equation}
e(n,t,U)=-\frac{2t\beta(U/t)}{\pi} \sin \left(\frac{\pi n} {\beta(U/t)}   \right)
\label{eq:e_BA}
\end{equation}
as the functional form for the energy per site when $n\leq 1$.
Here $\beta(U/t)$ is found, by using a Newton-Raphson procedure, so that the three limits are satisfied.  
The exchange-correlation functional for the BA-LDA\cite{LimaPRL2003}  is 
\begin{equation}
E_{xc}^{BA-LDA}=\sum_{i} e_{xc}(n_{i},t,U)
\label{eq:ExcBALDA}
\end{equation}
where, analogously to standard DFT, a local functional is created by subtracting the per site non-interacting kinetic energy and Hartree energy for the homogeneous Hubbard model 
\begin{eqnarray}
\nonumber e_{xc}(n_{i},t,U)&=&e(n_{i},t,U)-e(n_{i},t,0)-e_{H}(n_{i},U) \\
\nonumber &=&-\frac{2t\beta(U/t)}{\pi} \sin \left(\frac{\pi n} {\beta(U/t)}   \right)\\
&+&\frac{4t}{\pi}\sin \left(\frac{\pi n_{i}}{2} \right)-\frac{Un_{i}^{2}}{4}.
\end{eqnarray}
Other choices for $e_{H}(n_{i},U)$ are possible but this one is often employed by considering that for balanced spins $\left\langle \hat{n}_{i,\sigma} \right\rangle= n_{i}/2$.\cite{Capelle13}
For $n_{i}>1$ the particle-hole transformation for the Hubbard model gives $e(n_{i}>1,t,U)= e(2-n_{i},t,U)+U(n_{i}-1)$ which for all $n_{i}$ values can be succinctly accounted for by using $e_{xc}(1-|n_{i}-1|,t,U)$.\cite{Akande10} This means that $v_{xc}^{BA-LDA}(n_{i}>1,t,U)=-v_{xc}^{BA-LDA}(2-n_{i},t,U)$
and there is a discontinuity at $n_{i}=1$ which, as noted in Ref.~\onlinecite{Akande10}, can cause convergence issues when self-consistently solving the KS equation in this case.

\subsection{Quantum Chemistry Lattice Density-Functional Theory}

For an orthonormal basis set of single-particle orbitals, the quantum chemistry Hamiltonian can be written as a lattice Hamiltonian using the notation of second quantization\cite{Szabo89} as
\begin{equation}
\hat{H}=\sum_{pq\sigma} h_{pq} \hat{a}_{p\sigma}^{\dagger} \hat{a}_{q\sigma} +\frac{1}{2} \sum_{pqrs\sigma\sigma'} \braket{pr}{qs} \hat{a}_{p\sigma}^{\dagger}\hat{a}_{r\sigma'}^{\dagger}\hat{a}_{s\sigma'}\hat{a}_{q\sigma},
\label{eq:QC_H}
\end{equation}
and it is this mapping that allows the powerful approach of DMRG to be used for quantum chemistry. As orbitals are mapped to sites then if we can calculate the exact site occupation we will have found the orbital occupation for the FCI wavefunction. Here $\sigma$ and $\sigma'$ label the spins, $h_{pq}$ are the one-electron integrals for spatial orbitals $\phi_{p}$ and $\phi_{q}$, while the two-electron integrals are

\begin{equation}
\braket{pr}{qs}=\int \int \frac{\phi_{p}^{*}(\vec{r}_{1})\phi_{r}^{*}(\vec{r}_{2}) \phi_{q}(\vec{r}_{1})\phi_{s}(\vec{r}_{2})}{\left| \vec{r}_{1}-\vec{r}_{2} \right|}  d\vec{r}_{1} d\vec{r}_{2} 
\end{equation}
and atomic units are used. For $K$ basis functions, the number of two-electron integrals will scale as $O(K^{4})$ but the evaluation of them for gaussian basis sets is fast so this would only become a bottleneck for very 
large basis sets. In this case, by localizing orbitals, approximations could be employed to only consider near orbitals and reduce the severity of this scaling.

 To create QC-LDFT we make Eq.~\ref{eq:QC_H} amenable to the construction of a L-DFT KS equation by using the fermionic anticommutation relations $\{\hat{a}_{i\sigma},\hat{a}_{j\sigma'}\}=0$,  $\{\hat{a}_{i\sigma}^{\dagger},\hat{a}_{j\sigma'}^{\dagger}\}=0$   
and $\{\hat{a}_{i\sigma}^{\dagger},\hat{a}_{j\sigma'}\}=\delta_{ij}\delta_{\sigma\sigma'}$ to give terms involving the site density (orbital occupation) operator $\hat{n}_{i}=\hat{a}^{\dagger}_{i}\hat{a}_{i}$.

\subsubsection{Reduced Hamiltonian}
 We first consider the two-electron terms $\braket{pp}{pp}$ and  $\sigma\neq \sigma'$ which would correspond to on-site repulsion in the Hubbard model if all the $\braket{pp}{pp}$ were set to $U$. Using the anticommutation relations we have
$\hat{a}_{p\sigma}^{\dagger}\hat{a}_{p\sigma'}^{\dagger}\hat{a}_{p\sigma'}\hat{a}_{p\sigma}=-\hat{a}_{p\sigma}^{\dagger}\hat{a}_{p\sigma'}^{\dagger}\hat{a}_{p\sigma}\hat{a}_{p\sigma'}=\hat{a}_{p\sigma}^{\dagger}\hat{a}_{p\sigma}\hat{a}_{p\sigma'}^{\dagger}\hat{a}_{p\sigma'}= \hat{n}_{p\sigma}\hat{n}_{p\sigma'}$. This means that the terms in the Hamiltonian can be written as $\frac{1}{2}\sum_{p\sigma\sigma'} \braket{pp}{pp} \hat{n}_{p\sigma}\hat{n}_{p\sigma'}$ which becomes 
$ \frac{1}{2}\sum_{p} \braket{pp}{pp} (\hat{n}_{p\uparrow}\hat{n}_{p\downarrow}+\hat{n}_{p\downarrow}\hat{n}_{p\uparrow})$ when summing over spins.

We then employ the approach from L-DFT\cite{Capelle13} of using $\left\langle \hat{n}_{i,\sigma} \right\rangle= n_{i}/2$ when the spins are balanced to give
\begin{equation}
E_{H}=\frac{1}{4}\sum_{p} \braket{pp}{pp} n_{p}^{2}.
\label{eq:E_H}
\end{equation}
The contribution to $v_{eff,i}$ in the KS equation is then
\begin{equation}
v_{H,i}=\frac{\partial E_{H}}{\partial n_{i}}=\frac{1}{2}  \braket{ii}{ii} n_{i}
\end{equation}
For clarity we note that $n_{i}$ the site density or occupation is the orbital occupation for the basis of orbitals that were mapped to the sites in the lattice Hamiltonian (Eq.~\ref{eq:QC_H}). It is not the occupation of the eigenfunctions of the KS L-DFT Hamiltonian in its single determinant wavefunction as by construction their occupation cannot be fractional.

We next have the $\braket{pr}{pr}$ terms where $r\neq p$ and the anticommutation relations now lead to $\frac{1}{2}\sum_{p,r,(p \neq r)\sigma\sigma'} \braket{pr}{pr} \hat{n}_{p\sigma}  \hat{n}_{r\sigma'}$. Using $\left\langle \hat{n}_{i,\sigma} \right\rangle= n_{i}/2$ and that there are four spin combinations gives another energy contribution in terms of the density that we denote as the second Hartree term (H2)
\begin{equation}
E_{H2}=\frac{1}{2}\sum_{p,r,(p \neq r)} \braket{pr}{pr} n_{p} n_{r}.
\end{equation}
Resulting in another contribution to $v_{eff,i}$ of
\begin{equation}
v_{H2,i}=\frac{\partial E_{H2}}{\partial n_{i}}= \sum_{r,(r\neq i)}  \braket{ir}{ir} n_{r}
\end{equation}
where we have used that $\braket{ir}{ir}=\braket{ri}{ri}$. 

At this point we can write a KS L-DFT equation for a reduced quantum chemistry Hamiltonian where two-electron integrals beyond $\braket{pp}{pp}$  and $\braket{pr}{pr}$ are neglected
\begin{eqnarray}
\nonumber \hat{H}_{red,KS}&=&\sum_{p,q,(p\neq q)} h_{pq} \hat{a}_{p}^{\dagger} \hat{a}_{q}\\
&+&\sum_{i} \left( v_{ext,i}+v_{H,i}+v_{H2,i}+v_{xc,i} \right) \hat{n}_{i}. 
\label{eq:redKS}
\end{eqnarray}
Here the first term is $\hat{T}$ while $v_{ext,i}=h_{ii}$, $v_{H,i}=\frac{1}{2}\braket{ii}{ii} n_{i}$ and $v_{H2,i}=\sum_{r,(r\neq i)}  \braket{ir}{ir} n_{r}$.

\subsubsection{Full Hamiltonian}

We now consider  terms of the form $\braket{pr}{ps}$ where $r\neq s$ and the anticommutation relations give $\hat{a}_{p\sigma}^{\dagger}\hat{a}_{r\sigma'}^{\dagger}\hat{a}_{s\sigma'}\hat{a}_{p\sigma}=-\hat{a}_{r\sigma'}^{\dagger}\hat{a}_{p\sigma}^{\dagger}\hat{a}_{s\sigma'}\hat{a}_{p\sigma}
=\hat{a}_{r\sigma'}^{\dagger}\hat{a}_{s\sigma'}\hat{a}_{p\sigma}^{\dagger}\hat{a}_{p\sigma}-\hat{a}_{r\sigma'}^{\dagger} \delta_{sp} \delta_{\sigma\sigma'}\hat{a}_{p\sigma}$. So for the contribution to the
quantum chemistry Hamiltonian we have
\begin{eqnarray}
\nonumber \frac{1}{2} \sum_{p,r,s,(r\neq s),\sigma\sigma'} \braket{pr}{ps} \hat{a}_{p\sigma}^{\dagger}\hat{a}_{r\sigma'}^{\dagger}\hat{a}_{s\sigma'}\hat{a}_{p\sigma} &=&\\
\nonumber \frac{1}{2}\sum_{p,r,s,(r\neq s),\sigma\sigma'} \braket{pr}{ps} \hat{a}_{r\sigma'}^{\dagger}\hat{a}_{s\sigma'}\hat{n}_{p\sigma}\\
-\frac{1}{2}\sum_{r,s,(r\neq s),\sigma} \braket{sr}{ss} \hat{a}_{r\sigma}^{\dagger}\hat{a}_{s\sigma}.
\end{eqnarray}
This does not have an expression in terms of only the site density, but when the spins are balanced we again use $\left\langle \hat{n}_{i,\sigma} \right\rangle= n_{i}/2$ to give the contribution to the KS equation. We take into account the sum over spins for the site density to give $\frac{1}{2}\sum_{p,r,s,(r\neq s)} \braket{pr}{ps} \hat{a}_{r}^{\dagger}\hat{a}_{s}n_{p}-\frac{1}{2}\sum_{r,s,(r\neq s)} \braket{sr}{ss} \hat{a}_{r}^{\dagger}\hat{a}_{s}$ which due to the occurrence of $\hat{a}^{\dagger}_{p} \hat{a}_{q}$ terms becomes an addition
to $\hat{T}$ in the KS equation.

A similar procedure for the contribution of $\braket{pr}{qr}$ where $p\neq q$ results in the terms $\frac{1}{2}\sum_{p,q,r,(p\neq q)} \braket{pr}{qr} \hat{a}_{p}^{\dagger}\hat{a}_{q} {n}_{r} -\frac{1}{2}\sum_{p,q,(p\neq q)} \braket{pq}{qq} \hat{a}_{p}^{\dagger}\hat{a}_{q}$ being included in the KS Hamiltonian as an addition to $\hat{T}$.

Finally the only remaining integrals to consider are $\braket{pr}{qs}$ where $p\neq q$ and $r \neq s$. After rearranging the creation and annihilation operators we have $\hat{a}_{p\sigma}^{\dagger}\hat{a}_{q\sigma}\hat{a}_{r\sigma'}^{\dagger}\hat{a}_{s\sigma'}-\hat{a}_{p\sigma}^{\dagger} \delta_{rq} \delta_{\sigma\sigma'}\hat{a}_{s\sigma'}$.  The first term cannot be rewritten using the site density and is a pure two-electron term so does not occur in the KS equation. This leaves $-\frac{1}{2}\sum_{p,q,s,(p\neq q, q \neq s) } \braket{pq}{qs} \hat{a}_{p}^{\dagger}\hat{a}_{s}$ as the addition to $\hat{T}$ when $p \neq s$ and to $v_{eff}$ when $p=s$.  

Combining these results with the reduced KS Hamiltonian (Eq.~\ref{eq:redKS}) gives the full KS Hamiltonian for QC-LDFT 

\begin{equation}
\hat{H}_{full,KS}=\hat{T}+\sum_{i} v_{eff,i} \hat{n}_{i}
\end{equation}
where
\begin{eqnarray}
\nonumber \hat{T}&=&\sum_{p,q,(p\neq q)} h_{pq} \hat{a}_{p}^{\dagger} \hat{a}_{q}+\frac{1}{2}\sum_{p,r,s,(r\neq s)} \braket{pr}{ps} \hat{a}_{r}^{\dagger}\hat{a}_{s}n_{p}\\
\nonumber &-&\frac{1}{2}\sum_{r,s,(r\neq s)} \braket{rr}{rs} \hat{a}_{r}^{\dagger}\hat{a}_{s}+\frac{1}{2}\sum_{p,q,r,(p\neq q)} \braket{pr}{qr} \hat{a}_{p}^{\dagger}\hat{a}_{q}n_{r} \\
\nonumber &-&\frac{1}{2}\sum_{p,q,(p\neq q)} \braket{pq}{qq} \hat{a}_{p}^{\dagger}\hat{a}_{q} \\
&-&\frac{1}{2}\sum_{p,q,s,(p\neq q, q \neq s, p \neq s) } \braket{pq}{qs} \hat{a}_{p}^{\dagger}\hat{a}_{s}     
\end{eqnarray}
and 
\begin{eqnarray}
v_{eff,i}=v_{ext,i}+v_{H,i}+v_{H2,i}+v_{H3,i}+v_{xc,i}.
\label{eq:veffQC}
\end{eqnarray}
Here the contributions to  $v_{eff,i}$ are the same as for the reduced Hamiltonian (Eq.~\ref{eq:redKS}) except there is now a third Hartree potential
\begin{equation}
v_{H3,i}=-\frac{1}{2}\sum_{q, (q\neq i) } \braket{iq}{qi}. 
\end{equation}

\subsection{Numerical Procedure}

For the BA-LDA in QC-LDFT we use $e_{xc}(n_{i},t_{i},U_{i})$ in the approximation for $E_{xc}$ (Eq.~\ref{eq:ExcBALDA}) as approximate $U$ and $t$ values are now site dependent. Through comparison of the quantum chemistry Hamiltonian (Eq.~\ref{eq:QC_H}) with that of the Hubbard model (Eq.~\ref{eq:HubbardModelwithVext}) we see that $U_{i}=\braket{ii}{ii}$ and take the average of the one-electron `hopping' integrals to calculate $t_{i}$ values 
\begin{equation}
t_{i}=-\frac{1}{2}\left( h_{i,i+1}+h_{i-1,i} \right). 
\end{equation}
We have periodic boundary conditions as all orbitals can, in principle, interact so that for $K$ orbitals $t_{K}=-\frac{1}{2}\left( h_{K,1}+h_{K-1,K} \right)$ and $t_{1}=-\frac{1}{2}\left( h_{1,2}+h_{K,1} \right)$. We generate the one-electron and two-electron quantum chemistry integrals using the program Molpro.\cite{MOLPRO_brief2012} 

As the BA-LDA is exact for three limiting cases of the homogeneous Hubbard model then it would be expected to work best when site densities or occupations are not too different from one another. In this case if substantially more orbitals are used than electrons then the chance that some $n_{i}$ are close to one is also reduced. Furthermore we would like the $t_{i}$ values to be non-negligible and similar, 
as if some are close to zero then the calculation of $\beta(U_{i}/t_{i})$ in the BA-LDA expression for the energy per site (Eq.~\ref{eq:e_BA}) will become unreliable. 
To make it more likely that the $t_{i}$ values are similar and that the site densities or occupations are not too far from homogeneity, we do not use Hartree-Fock molecular orbitals but begin with $K$ atomic orbitals then orthogonalize them in a balanced way by using symmetric orthogonalization.\cite{Lowdin56} This transforms the non-orthogonal orbitals $\int \phi_{r}^{*} \phi_{s} =\Delta_{rs}$ to $\int \tilde{\phi}_{u}^{*} \tilde{\phi}_{v} =\delta_{uv}$
using $\tilde{\phi}_{u}=\sum_{i} \Delta^{-\frac{1}{2}}_{iu} \phi_{i}$.

With the aim of making the self-consistent calculations more robust and accelerating convergence we use a Newton-Raphson approach to solve 
\begin{equation}
\vec{G}(\vec{n})=\vec{O}(\vec{n})-\vec{n}=0.
\end{equation} 
Here $\vec{O}(\vec{n})$ is the site density from the KS eigenfunctions when the site density $\vec{n}$ is used in the KS equation. This gives the site density for iteration $i+1$ as
 \begin{equation}
\vec{n}(i+1)=\vec{n}(i)-\bm{J}^{-1}\vec{G}(\vec{n}(i))
\end{equation}
where $\bm{J}$ is the Jacobian matrix for $\vec{G}$.  We calculate $\bm{J}$ numerically with a step size of $0.01$ as solving the KS equation for given site densities is very fast.  To also improve stability we implement density mixing where $ \vec{n}(i+1)_{mix}= 0.2\vec{n}(i+1)+0.8\vec{n}(i)$.  From the fourth iteration we check the average difference between the input and output site densities for the KS equation which we denote as the error
\begin{equation}
\text{Error}=\frac{1}{K} \sum_{i=1}^{K} \left| O_{i}(\vec{n})-n_{i} \right|. 
\end{equation}
We use a threshold of $10^{-7}$ for this to ascertain if convergence has been reached when solving the KS equation self-consistently.

\section{Results}

We demonstrate QC-LDFT with the BA-LDA by calculating potential energy curves as the bond length is varied for a linear chain of six hydrogen atoms. The 3-21G basis set is employed resulting in 12 single-particle orbitals. We use a default ordering for the orbitals in the lattice
so sites $1$ and $2$ represent the symmetrically orthogonalized atomic orbitals of the first hydrogen, sites $3$ and $4$ represent those of the second hydrogen and so on.
  
\begin{figure}[h!]\centering
\includegraphics[width=.45\textwidth]{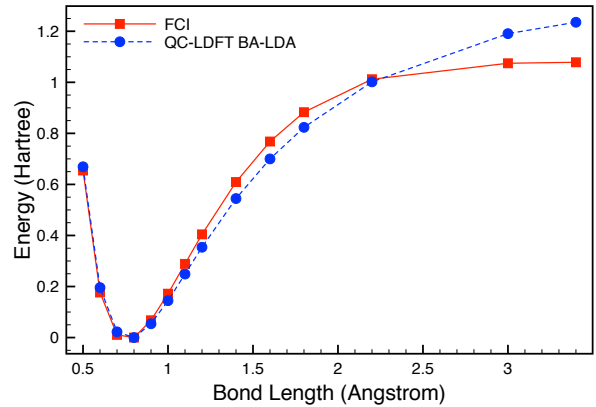}
\caption{Reduced Hamiltonian energy results from FCI and QC-LDFT with the BA-LDA for a linear chain of six hydrogens as the bond lengths are varied, using the 3-21G basis and shifting the potential curves
so that both have their minimum at zero.}\label{Fig:H6reduced}
\end{figure}

First we investigate the reduced Hamiltonian when two-electron integrals beyond $\braket{pp}{pp}$  and $\braket{pr}{pr}$ are neglected and its corresponding L-DFT KS equation (Eq.~\ref{eq:redKS}). 
 We see in Fig.~\ref{Fig:H6reduced} that a binding curve is recovered by the FCI results despite using a reduced Hamiltonian.  This fits in with results\cite{Chiappe07} that a Hubbard model with intersite repulsion could have
parameters derived to reasonably describe potential curves of the hydrogen molecule.  The FCI and QC-LDFT potential curves are shifted in Fig.~\ref{Fig:H6reduced} so both have zero as their minimum and we see that QC-LDFT
reproduces the shape of the curve and is in good agreement with FCI except around bond lengths of 3\AA~and greater where the QC-LDFT results are a little high. 

We found that $48400$ determinants were needed for the FCI calculation and due to the use of symmetric orthonormalization of
the atomic orbitals then even at the equilibrium bond length of 0.8\AA~very many determinants are important. We quantify this using an indicator\cite{MCCImetaldimers,MRinQC} of the FCI wavefunction's multireference character
\begin{equation} 
MR=\sum_{i} |c_{i}|^{2}-|c_{i}|^{4}
\end{equation}
where $c_{i}$ is the coefficient of determinant $i$ and the wavefunction is normalized so that $\sum_{i} |c_{i}|^{2}=1$. $MR$ is zero for a wavefunction consisting of a single determinant while the value approaches one as the number of important determinants increases. Even at the equilibrium bond length we find $MR=0.9999$ which demonstrates the very strong multireference character when using atomic orbitals with symmetric orthonormalization.

\begin{figure}[h!]\centering
\includegraphics[width=.45\textwidth]{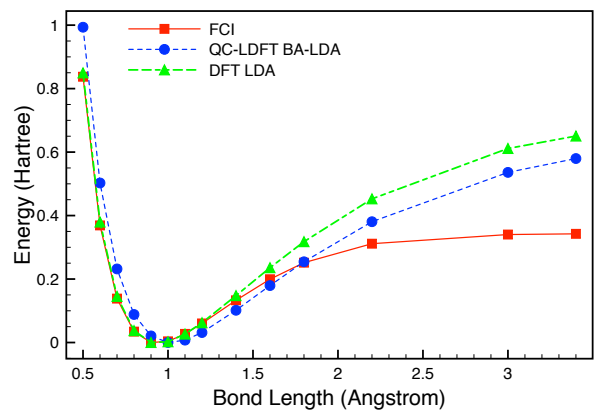}
\caption{Full Hamiltonian energy results from FCI, QC-LDFT with the BA-LDA, and standard DFT with the LDA for a linear chain of six hydrogens as the bond lengths are varied, using the 3-21G basis and shifting the potential curves
so that all have their minimum at zero.}\label{Fig:H6Full}
\end{figure}

The full quantum chemistry Hamiltonian is now considered using FCI, and QC-LDFT with the BA-LDA. For comparison, results from standard DFT with the LDA are calculated using Molpro.\cite{MOLPRO_brief2012} We see in Fig.~\ref{Fig:H6Full} that the general shape of the FCI binding curve is again captured by QC-LDFT but the discrepancy at large bond lengths is more apparent.

 We speculate that this is due to the BA-LDA being based on the Hubbard model which means it does not include the one-electron integrals beyond nearest neighbours nor the extra Hartree terms that occur in the quantum chemistry KS Hamiltonian (Eq.~\ref{eq:veffQC}). This fits in with the difference with the exact result being less pronounced for the reduced Hamiltonian in Fig.~\ref{Fig:H6reduced} as there is is only one additional Hartree term for the KS equation in this case (Eq.~\ref{eq:redKS}). The minimum energy is at 0.9\AA~for both FCI and DFT, while QC-LDFT is close to this at 1.0\AA.  The DFT results are closer to FCI at shorter bond lengths but QC-LDFT performs better as the bonds are elongated.

\begin{figure}[h!]\centering
\includegraphics[width=.45\textwidth]{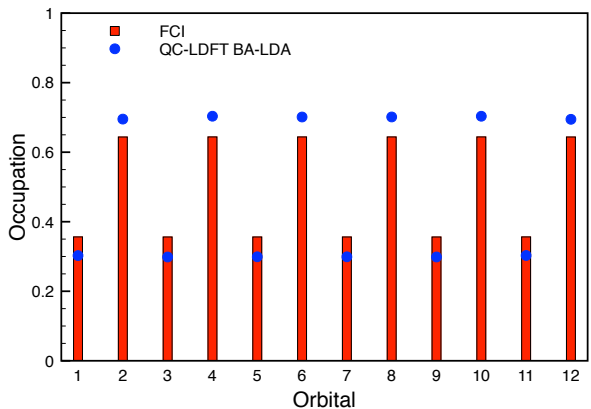}
\caption{Full Hamiltonian orbital occupancy results for FCI and QC-LDFT with the BA-LDA for a linear chain of six hydrogens at a bond length of 3.4\AA~using the 3-21G basis.}\label{FigH6_3_4occ}
\end{figure}

 We quantify the overall error in the potential curves compared with FCI using $\sigma_{\Delta E}$ from Refs.~\onlinecite{MCCInatorb,LDFT18} where
\begin{equation}
\sigma_{\Delta E}=\sqrt{\frac{1}{d}\sum_{j=1}^{d}(\Delta E_{j}-\mu_{\Delta E})^{2}}
\end{equation}
is the standard deviation of the difference in energies $\Delta E_{j}=E_{j}^{FCI}-E_{j}^{approx}$ for all $d$ points in the potential energy curve and $\mu_{\Delta E}$ is the mean value of $\Delta E$.
This takes into account all of the points and that the curves can be shifted by a constant.  This gives $0.086$ and $0.100$ Hartree for QC-LDFT and DFT respectively, showing that for these points QC-LDFT is slightly more accurate by this measure.
Again the  orbitals used means that the problem is strongly multireference for FCI and QC-LDFT at all points considered. In addition the values for $U_{i}/t_{i}$ are around $1$ to $5$ at both 1.0 \AA~ and 3.4 \AA. We see in Fig.~\ref{FigH6_3_4occ} that at 3.4 \AA, when there is a more noticeable difference between the FCI and QC-LDFT potential energy curves, the orbital occupations calculated using QC-LDFT are slightly different to the FCI results but have a very similar pattern. This suggests that a functional designed specifically for QC-LDFT should be able to correct the discrepancy in energies for this region.

\section{Summary}

We created a lattice density-functional theory for {\it ab initio} quantum chemistry or physics (QC-LDFT) by considering the quantum chemistry Hamiltonian in the notation of second quantization where orbitals are mapped to sites on a lattice and deriving its L-DFT Kohn-Sham equation. This represents an efficient approach to approximate the energy and orbital occupation of the full configuration interaction wavefunction as for $K$ basis functions then the cost of solving the L-DFT Kohn-Sham equation scales as $O(K^{3})$. We demonstrated QC-LDFT on a linear chain of six hydrogen atoms with a basis set of twelve orbitals as the bond length was varied and tested the approximate BA-LDA\cite{LimaPRL2003} functional for this case. Remarkably, despite using this approximate functional designed for the Hubbard model,
QC-LDFT captured the shape of the FCI potential energy curves for both a reduced and full Hamiltonian. In the latter case a discrepancy was more noticeable at stretched bond lengths however there was an improvement over standard DFT here. Future work will
consider optimizing the orbital ordering in the lattice, smoothing\cite{Karlsson11} of $v_{xc}$ around $n=1$ and going beyond the BA-LDA functional so that QC-LDFT can be applied successfully to more complex multireference or strongly-correlated molecules.

\section*{Acknowledgements}
JPC thanks the EPSRC for support via the platform grant EP/P001459/1.

\providecommand{\noopsort}[1]{}\providecommand{\singleletter}[1]{#1}%

\end{document}